\PassOptionsToPackage{table}{xcolor}

\documentclass[runningheads]{llncs}
\usepackage{tabularx}
\usepackage[dvipsnames]{xcolor}
\usepackage{amsmath}
\usepackage{booktabs}
\usepackage{tcolorbox}
\newcommand{\robertaclaimnoev}{\textsc{roberta-mnli-claim-no-ev}}

\newcommand{\claimonly}{\textsc{Claim-Only}}
\newcommand{\oracleq}{\textsc{FCDecomp}}
\newcommand{\oracle}{\textsc{Oracle}}
\newcommand{\claimdecomp}{\textsc{ClaimDecomp}}
\newcommand{\pgmfc}{\textsc{PgmFC}}
\newcommand{\qgen}{\textsc{QGen}}

\usepackage{caption}
\usepackage{subcaption}

\newcommand{\combmaxn}{\textsc{CombMax-Norm}}

\newcommand{\gptt}{\textsc{gpt-3.5-turbo}}

\newcommand{\gpttgold}{\textsc{gpt-3.5-turbo-gold}}

\newcommand{\robertaqtempplus}{\textsc{roberta-mnli-qtemp++}}
\newcommand{\robertagold}{\textsc{roberta-mnli-gold}}
\newcommand{\robertamnlilarge}{\textsc{roberta-large-mnli}}
\newcommand{\naive}{\textsc{Naive-Classifier}}

\usepackage{enumitem}
\usepackage{amssymb}% http://ctan.org/pkg/amssymb
\usepackage{pifont}% http://ctan.org/pkg/pifont
\newcommand{\cmark}{\ding{51}}% check mark
\newcommand{\xmark}{\ding{55}}% cross mark

\usepackage{xcolor} % colors
\newcommand{\quantempplus}{\textsc{QuanTemp++}}

\newcommand{\name}{\textsc{FCDecomp}}

\usepackage[table]{xcolor}
\definecolor{GREEN}{RGB}{84,130,53}

\usepackage{booktabs}
\usepackage[T1]{fontenc}
\usepackage{graphicx}
\begin{document}
\title{ A Benchmark for Open-Domain Numerical Fact-Checking  Enhanced by Claim Decomposition }
\author{Venktesh V\inst{1}\orcidID{0000-0001-5885-2175} \and
Deepali Prabhu\inst{2} \and
Avishek Anand\inst{3}\orcidID{0000-0002-0163-0739}}
\authorrunning{Venktesh V et al.}
\institute{Stockholm University, Sweden \email{venktesh.viswanathan@dsv.su.se}\\ \and
Independent Researcher \\
% \url{http://www.springer.com/gp/computer-science/lncs} 
\and
Delft University of Technology, Netherlands,
\email{Avishek.Anand@tudelft.nl}}
\maketitle              % typeset the header of the contribution
\begin{abstract}
Fact-checking numerical claims is critical as the presence of numbers provide mirage of veracity despite being fake potentially causing catastrophic impacts on society. The prior works in automatic fact verification do not primarily focus on natural numerical claims. A typical human fact-checker first retrieves relevant evidence addressing the different numerical aspects of the claim and then reasons about them to predict the veracity of the claim. Hence, the search process of a human fact-checker is a crucial skill that forms the foundation of the verification process. Emulating a real-world setting is essential to aid in the development of automated methods that encompass such skills. However, existing benchmarks employ heuristic claim decomposition approaches augmented with weakly supervised web search to collect evidences for verifying claims. This sometimes results in less relevant evidences and noisy sources with temporal leakage rendering a less realistic retrieval setting for claim verification. Hence, we introduce QuanTemp++: a dataset consisting of natural numerical claims, an open domain corpus, with the corresponding relevant evidence for each claim. The evidences are collected through a claim decomposition process approximately emulating the approach of human fact-checker and veracity labels ensuring there is no temporal leakage. Given this dataset, we also characterize the retrieval performance of key claim decomposition paradigms. Finally, we observe their effect on the outcome of the verification pipeline and draw insights. The \emph{code for data pipeline along with link to data} can be found at
 \url{https://github.com/VenkteshV/QuanTemp\_Plus}

\end{abstract}

\section{Introduction}
\begin{table}[ht]
\centering
\resizebox{\linewidth}{!}{
\begin{tabular}{lcccccccc}
\toprule
Paper & \#Claims & Claim Src & Natural? & Evidence Src & \multicolumn{2}{c}{No Leakage} & \(\text{Numerical}^{*}\) \\
\cmidrule(lr){6-7}
& & & & & Gold & Temporal & \\
\midrule
Thorne and Vlachos\cite{thorne-vlachos-2017-extensible} & 7k & KB & \xmark & KB & NA & NA & \xmark \\
Vlachos and Ridel\cite{vlachos-riedel-2015-identification} & 7k & KB & \xmark & KB & NA & NA & \cmark \\
FavIQ\cite{FAVIQ} & 188k & QA & \xmark & Wikipedia & NA & NA & \xmark \\
VitaminC\cite{VitaminC} & 326k & Wikipedia & \xmark & Wikipedia & NA & NA & \xmark \\
HOVER\cite{jiang2020hover} & 26k & QA & \xmark & Wikipedia & NA & NA & \xmark \\
SciFact\cite{scifact} & 1.4k & Science Articles & \xmark & Science Articles & NA & NA & \xmark \\
SciFactOpen\cite{wadden-etal-2022-scifact} & 279 & Science Articles & \xmark & Science Articles & NA & NA & \xmark \\
WICE\cite{WICE} & 2k & Wikipedia & \xmark & Wikipedia & NA & NA & \xmark \\
FEVER\cite{FEVER} & 185k & Wikipedia & \xmark & Wikipedia & NA & NA & \xmark \\
FEVEROUS\cite{FEVEROUS} & 87k & Wikipedia & \xmark & Wikipedia & NA & NA & \xmark \\
\bottomrule
MultiFC\cite{multifc} & 36k & FCW & \cmark & Open Web & \xmark & \xmark & \xmark \\
ClaimDecomp\cite{claimdecomp} & 1.2k & FCW & \cmark & Fact Check Articles & NA & NA & \xmark \\
FinFact\cite{finfact} & 3.4k & FCW & \cmark & Open Web & \xmark & \xmark & \xmark \\
WatClaimCheck\cite{watclaimcheck} & 34k & FCW & \cmark & Open Web & \cmark & \cmark & \xmark \\
LIAR\cite{liar} & 13k & FCW & \cmark & NA & NA & NA & \xmark \\
QABriefs\cite{qabriefs} & 8.8k & FCW & \cmark & Open web & \cmark & \xmark & \xmark \\
AviriTec\cite{averitec} & 4.5k & FCW & \cmark & Open Web & \cmark & \cmark & \xmark \\
Jandhagi and Pujara\cite{QVS} & 500 & News Articles & \cmark & Select Datasets & \cmark & \cmark & \cmark \\
COVIDFACT\cite{saakyan-etal-2021-covid} & 4.1k & Reddit & \cmark & Open Web & \cmark & \cmark & \xmark \\
QuanTemp\cite{quantemp} & 15k & FCW & \cmark & Open Web & \cmark & \xmark & \cmark \\
\rowcolor{blue!25}
\quantempplus{}(\textbf{ours}) & 15k & FCW & \cmark & Open Web & \cmark & \cmark & \cmark \\
\bottomrule
\end{tabular}
}
\caption{Comparison of \quantempplus{} to other related fact checking datasets. Leakage refers to either the presence of fact-checking articles (Gold) in the evidence corpus or the inclusion of evidence items published after the corresponding claim (Temporal). We also judge if a significant number of \(\text{Numerical}^{*}\) claims are present in the dataset. In the table, FCW refers to Fact-Checking Websites. }
\label{tab:dataset_all}
\vspace{-2em}
\end{table}

The spread of misinformation and disinformation has surged with the rise of advanced digital tools to disseminate information. Although manual verification of such claims are performed by experts, the sheer volume of misinformation motivates the need for automated fact-checking approaches\cite{cohen2011computation}. 

Of particular significance among the wide range of claims are numerical claims which comprise quantitative or temporal information \cite{quantemp}. These claims are significant because of the ``illusion of numeric truth effect" where claims with quantitative information provide a mirage of truth \cite{sagara2009consumer} potentially leading to catastrophic effects. For instance, unsubstantiated incomplete claims were made by Purdue Pharma to market their drug which is believed to be the kickstarter of the Opioid pandemic in America, resulting in the loss of over 500,000 lives to date\cite{van2009promotion}. Verification of numerical claims requires numerical reasoning ability which involves understanding numeric patterns and mathematical concepts like arithmetic, numerical estimation, and data interpretation.  While several works focus on automated verification of natural claims \cite{programFC,shah2022numerical,vlachos-riedel-2015-identification,shah2023concord,QVS}, only a handful focus on claims that require processing numerical information. However, these techniques only focus on the detection of numerical claims\cite{shah2023concord,shah2022numerical} or are restricted to specific simple statistical properties\cite{thorne-vlachos-2017-extensible,vlachos-riedel-2015-identification}. QuanTemp \cite{quantemp} was the first benchmark to focus on diverse numerical claims but is limited in evidence quality due to temporal leakage.

A typical human fact-checker utilizes various skills to verify a natural numerical claim. A given numerical claim can have different explicit and implicit information needs that first needs to be addressed.  Therefore, relevant evidence is gathered from various sources, including the open web. This evidence is then used to reason about and determine the veracity of a natural numerical claim.  The search process of a human fact-checker which involves decomposing the claim along different aspects is a crucial skill that forms the foundation of the verification process. Emulating a real-world setting is essential to aid in the development of automated methods that encompass such skills.

\begin{figure}[h!]
    \includegraphics[width=\textwidth]{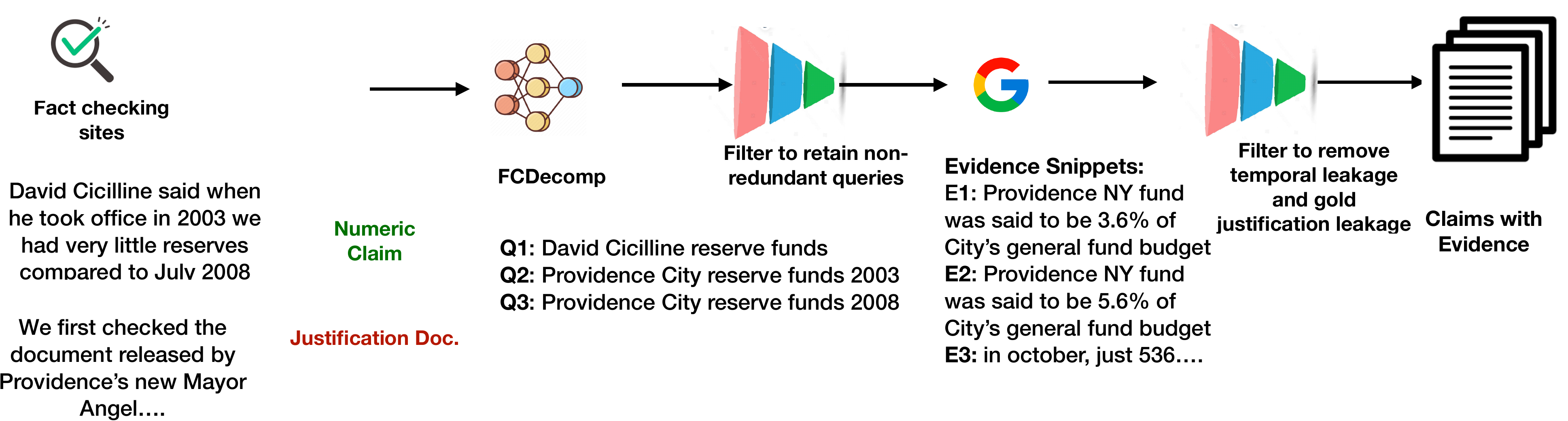}
    \caption{Data creation pipeline of \quantempplus{}}
    \label{fig:data-collection}
\end{figure}
\vspace{-1em}
Current datasets mainly feature synthetic claims with lexical biases \cite{FAVIQ,Chen2023ComplexCV}, while those with natural claims often suffer from gold or temporal leakage\cite{averitec} or lack focus on numerical claims (see Table 1.1). The presence of leaked fact verification articles in the corpus trivializes retrieval, while including evidence published after the claim makes the corpus unrealistic. Consequently, any automatic method designed under these conditions is likely to fail during real-time deployment. Hence, our goal is to provide a realistic dataset that addresses all these gaps, aiding the right development of automated methods for verifying natural numerical claims. 

Forming a large-scale dataset to address the above gaps through crowdsourcing can be challenging as obtaining manual annotations from expert fact-checkers is expensive\cite{expensive}. Therefore, we employ weak supervision to extend QuanTemp\cite{quantemp} and create
\textbf{\quantempplus{}}: a dataset consisting of about 15k natural numerical claims, an open domain corpus consisting of 165.7k records, and the corresponding evidence relevance and veracity labels of the claims. To improve the evidence quality collected from web, we propose \name{}, a claim decomposition that emulates the search process of human fact-checkers. Evidences collected using queries from \name{} helps in realistically evaluating and understanding retrieval and verification bottlenecks in automated fact-checking methods.

We aim to address the following research questions:
\begin{enumerate}
\setlength\itemsep{1em}
\item \textbf{RQ1}: Does \name{} based decomposition help retrieve quality evidence from the web for the verification of natural numerical claims?
    \item \textbf{RQ2}: How do existing claim decomposition methods perform in terms of evidence retrieval to verify numerical claims in \quantempplus{}?
    \item \textbf{RQ3}: What is the downstream impact of these claim decomposition methods on the task of verification of numerical claims?

\end{enumerate}

\section{Related Work}

The automated fact-checking process entails a three-stage pipeline namely claim detection, evidence retrieval and claim-verification. In this section, we discuss limitations of prior datasets and motivate the need for a realistic open-domain setup currently absent in existing datasets.

Majority of the existing datasets primarily focus on synthetically curated claims from Wikipedia articles \cite{jiang2020hover,thorne-etal-2018-fever,aly2021feverous,creditassess,schuster2021vitamin,sathe-etal-2020-automated}. Evaluation of automated fact-checking systems on such benchmarks is not representative of real-world claims encountered by fact-checkers motivating the need for curation of such real-world claims.  Recent efforts like \cite{kamoi2023wice} focus on creating more realistic claims from Wikipedia by identifying cited statements, but fall short and do not reflect the typical distribution of claims verified by fact-checkers as the claims are still synthetic. More recently, substantial effort has been diverted to collection of previously fact-checked real-world claims in politics \cite{multifc,averitec,claimdecomp}, science \cite{wadden-etal-2020-fact,vladika2023scientific,wright2022generating}, health \cite{kotonya2020explainable} and climate \cite{diggelmann2021climatefever} domains.  However, none of these benchmarks primarily focus on numerical claims, which comprise a substantial portion of real-world claims occurring in political debates or other contexts which comprise numerical quantities. Hence, automated fact-checking systems must account for numerical understanding capabilities that are required for evidence retrieval and claim verification to fact-check such claims. Moreover, existing datasets also do not have a realistic evidence collection due to leakage of manual fact-checker ``gold" evidence documents and related verdicts. This leakage may induce the claim verification models to learn shortcuts or overfit to fact-verification justification from manual fact-checkers. Additionally, the data collection pipeline for these benchmarks rely on heuristic search from open web search engines which solely uses the claim as search query. While this may work well for simple synthetic claims, real-world claims are often complex that require multiple aspects to be verified. Hence, using only the original claims as query may limit the evidences collected due to reliance on surface level matches to claim ignoring the numerical and other aspects of the claim.

More recently, QuanTemp \cite{quantemp} addresses these gaps by curating a realistic open-domain benchmark focusing entirely on numerical claims. While QuanTemp addresses leakage of ``gold" evidence documents by filtering out fact-checking websites, it still suffers from \textit{temporal leakage} issue. This issue occurs when evidence snippets published after the date the claim to be verified was made are also considered as relevant evidence and included in the collection. While Averitec \cite{averitec} proposes to mitigate temporal leakage, it neither focuses on numerical claims nor on handling leakage of ``gold" evidence documents. To mitigate these issues, we handle both temporal and ``gold" evidence leakage. Additionally, to cover all aspects of claims, we propose a claim decomposition process which emulates the fact-checker search process. We achieve this by using their justification description as part of data collection pipeline with particular focus on generating queries around numerical aspects of the claims. This results in high quality relevant evidence compared to existing approaches.
\begin{tcolorbox}[title= \name{}s Prompt]
\small
\textcolor{teal}{[CLAIM]} \\
Prime Minister Narendra Modi breached the election protocol by addressing a rally in Howrah on April 6.
\textcolor{teal}{[END]}

\textcolor{orange}{[PASSAGE]}\\
Modi addressed a rally in Cooch Behar and Howrah’s Dumurjola on April 6, where polls were held on April 10. The silence period was not breached. ...... In Dumurjola, the voting took place in the fourth phase on April 10. Therefore, Modi did not breach the 48 hours silence period. The voting for West Bengal assembly elections for 294 seats is taking place from March 27 to April 29. \dots\\
Published: 2021-04-07. 
\textcolor{orange}{[END]} 

\textcolor{blue}{[QUERIES]}
1. Modi rally in Howrah 2021

     2. Prime Minister Narendra Modi rally on April 6
     
      3. election in Howrah 2021
      
       4. locations of Howrah voting
        
         5. silence period affects campaigning and media coverage of elections
\textcolor{blue}{[END]}

Using the above as an example, generate at most 10 independent, short, and diverse Google search phrases required to verify or debunk the below claim labeled under \textcolor{teal}{[CLAIM]}.
Note: Generate diverse questions tending to the different numerical and temporal aspects both implicit and explicit to the claim.
\\ Note: Use the passage under \textcolor{orange}{[PASSAGE]} for reference to generate queries. 
%\\ Do not generate queries for the passage.  
%\\ Note: You must exclude any questions about fact-checking. 
\end{tcolorbox}
\captionof{figure}{ Prompt used to generate sub-queries given claim and justification document in the data creation pipeline of \quantempplus{} \\}
\label{example:prompt_agq} 
\section{Data Construction Approach}

An overview of our data collection pipeline  is shown in Figure \ref{fig:data-collection}. We primarily focus on curating a realistic evidence collection without temporal leakage or ``gold" justification leakage.  Additionally to cover all aspects of the claim we propose a claim decomposition process that generates queries around different aspects of the claims by extracting search patterns from justification document of manual fact-checkers. We observe that these queries help in collecting relevant evidence published prior to the date the claim was made. Compared to prior evidence collection approaches \cite{programFC,claimdecomp,multifc}, it also helps reduce noisy evidence stemming from surface level matches to original claim and provides better coverage for different aspects of the claim.

\subsection{\name{}: Automatic Query Generation Emulating Fact-checker Search Process}
\label{fcdecomp}

Manual fact checkers often search multiple news sources or documents on the open web to fetch relevant evidence articles to verify claims. They perform this search by usually generating search queries around different aspects of the claim to ensure the collected evidences are sufficient to verify the complete claim. This search process though not explicitly mentioned on all fact-checking websites is implicit in the justification provided by fact-checkers for verdicts assigned to previously fact-checked claims.

To emulate their thought process, we use few-shot learning to generate these web search queries using the justification document provided for each claim. We hypothesize that utilizing a claim's justification document in a few-shot learning setup with a Large Language Model (LLM) would generate both explicit and implicit queries required to verify a natural numerical claim. 
% LLM’s, specifically Open AI’s GPT models have shown to be effective to simulate user search queries\cite{9999111,10.1145/3366423.3380193}, hence we use the gpt-3.5-turbo model to generate the questions.
The prompt to generate Web search queries includes the claim, its publication date, justification document, and one static example crafted manually by the authors. The instruction specifies the generator to produce queries that sufficiently address the \textbf{numerical aspects} of the claim. The full prompt can be found in the Figure.\ref{example:prompt_agq}. 

To mitigate hallucination and redundant queries, we formulate a filtering process that ensures only queries which are unique and relevant are retained. The filter is constructed using the Maximal Marginal Relevance (MMR) measure.
% Within the algorithm, we employ the paraphrase-MiniLM-L6-v2 model to assess the similarity between texts. 
Given that the queries generated may be explicit or implicit we need to ensure that the selected queries are still relevant but don’t deviate from the claim's core aspects. Therefore, we employ a weighted average approach, to combine the MMR scores with respect to the claim and justification document to form the final scores for each query. Finally, we apply a threshold to retain only the most relevant and independent queries for each claim. Hence, for a given claim \( C \), justification document \( J \) and their corresponding sub-queries \( q_1, \ldots, q_k \), the final set of queries \( Q \) is formed by:
\[
s_{q_i} = \alpha \cdot \text{MMR}(q_{i}, C,\lambda) + (1-\alpha) \cdot \text{MMR}(q_{i}, J,\lambda) ; \quad Q = \{ q_i \mid s_{q_i} > \gamma \text{ for } i \in \{1, k\} \}
\]

% s

Here, \(\alpha\)  is the coefficient used to control the importance needed to be given to relevance with respect to the claim versus that to the justification document, \(\lambda\) is the diversification factor used in MMR and \(\gamma\) is the threshold for selecting a query.

\name{} outputs a variable number of diverse, high-quality queries emulating a manual fact-checker's search process described in the \textbf{justification document} to relevant evidence per claim. The average number of queries per claim is 6.76, which is higher than approaches like PgmFC\cite{programFC} and ClaimDecomp\cite{claimdecomp}. 

\subsection{Collecting Evidence}

After the unique high quality queries are retained, we collect search results from public search engines. However, unlike prior works \cite{multifc,schlichtkrull2023averitec,claimdecomp} which do not account for evidence leakage from fact-checker websites, we filter out evidence from over 150 fact-checking domains adopting the list from \cite{quantemp}. This ensures that the justification written by fact-checkers for the claims does not leak into evidence collected to prevent fact-checking models from learning shortcuts.  While QuanTemp \cite{quantemp} also focuses on preventing such leakage, it does not account for \textit{temporal leakage} where  evidence snippets published after the date the claim was made. This does not reflect realistic fact-checking scenario as systems and manual fact-checkers would only have access to relevant evidence prior to when the claim is made.  To prevent temporal leakage, we add the \textbf{before: filter} to the query issued to the search engine APIs, which restricts the results to those published before the date provided. Then we also perform basic de-duplication to avoid persisting redundant evidences in the collection. The evidence snippets collected from web using these queries are mapped to the claim as relevant evidences after basic de-duplication and also retaining documents that have high diversity as measured by MMR, to form qrels. Since a realistic open-domain retrieval setup also comprises distractors in the corpus, we retain other noisy evidence snippets retrieved from the web prior to filtering to form qrels in the corpus

\subsection{Distilling \name{} Queries and Evidence Aggregation}

While, \name{} helps generate high quality queries, to collect evidence, the queries cannot be used at test time. This is primarily because they rely on justification document for fact-checkers which is not available at test-time in a  realistic automated fact-checking setup. Hence, we propose to train a Language Model to generate such queries by using the claim and corresponding queries generated for the training set using a LLM.

Given the original Claim $C$ from training set, we employ a LLM like \gptt{} to generate a structured list of sub-queries 
$Q = \{q_1, q_2, \ldots, q_n\}$ as described in Section \ref{fcdecomp}.

Since, fine-tuning and inference using another LLM for query generation would be expensive and infeasible, we train a relatively smaller language model $f_\theta$ (e.g., FLAN-T5-LARGE : 780M parameters) 
to imitate the LLM's decomposition behavior. 
The model takes as input the claim $c$ (optionally prefixed with an instruction, 
e.g., ``\textit{Decompose the following claim into search queries:}'') 
and is trained to output the corresponding list of queries $Q$.
The model is optimized using the standard next-token prediction (cross-entropy) loss:
\[
\mathcal{L}(\theta) = - \sum_{(c, Q) \in \mathcal{D}_{\text{train}}} 
\sum_{t=1}^{|Q|} \log P_\theta \big(y_t \mid y_{<t}, c \big),
\]
where $y_t$ is the $t$-th token in the target query list $Q$. At inference time, the model generates a list of sub-queries given the instruction and the claim. 
\vspace{-1em}
\subsection{ Evidence Aggregation for Claim Verification}
The existing claim decomposition approaches like Claimdecomp \cite{claimdecomp}, ProgramFC \cite{programFC} primarily generate queries or sub-claims from original claim followed by retrieval of top-k evidences per query/sub-claim. Then simple concatenation is employed to merge different retrieval lists for claim verification without considering relative ordering. Rank fusion approaches can help in a principled combination of evidences from queries/sub-claims \cite{irfusion,dark_horse_error,irfusion2}.  We evaluate different such approaches and compare them with heuristic top-1 and concatenation based methods for all baselines and \name{}. We observe that while CombSUM and CombMAX provide improvements over naive top-k or concatenation approaches, CombMAX-Norm which applies CombMAX to normalized relevance scores works best to merge evidences across queries from decomposed claims.  %Finally, for claim-verification we employ fine-tuned NLI models or few-shot prompted LLMs given the claim and aggregated evidence snippets.
\vspace{-1em}
\section{Experimental Setup}

\textbf{Dataset Statistics}: We source the claims from QuanTemp and detailed statistics can be found in \cite{quantemp}. It comprises 9935 training, 3084 validation and 2495 test claims. Our evidence collection approach results in 165.7k evidence records.

\textbf{Qualitative Analysis of Decomposed Queries and Evidences }:
To ensure that the queries resulting from \name{} and subsequent evidences collected are of good quality we perform qualitative analysis. Inspired by comprehensiveness measure proposed in \cite{claimdecomp}, we measure the completeness, redundancy and relevance of both the generated queries from \name{} and resulting evidence  with respect to the claim. These measures are defined as follows: 

\textbf{Completeness}: Here the annotators are primarily asked to rate on a scale of (1-5) whether the generated queries cover all aspects of the claim.

\textbf{Redundancy}: The queries generated should be concise and cover independent aspects of the claim (rating 1). if multiple queries cover the same aspect of the claim, they are redundant (rated 5). Depending on extent of redundancy, a rating between 1-5 is assigned. The annotators are also asked to judge the \textbf{Relevance} (0/1) of the generated queries to the claim. Annotators are also asked to rate evidences along the same dimensions. To assess inter-annotator agreement, we calculate Fleiss’ kappa \cite{fleiss}
for each of the annotation tasks. \\
\textbf{Quantitative Analysis of Data Quality}:

Since extensive manual annotation of whole evidence collection is expensive, to automatically evaluate the quality of evidence collected, we compute the upper bound on fact-verification performance attainable using the evidences. We then compare them with  performance attained using gold justification documents. 

\begin{table*}[h]
    \centering
    \begin{tabularx}{\textwidth} { 
       >{\centering\arraybackslash} l 
       c 
       X 
       X 
       X 
       X 
       X
       X 
 }
        \toprule
        \textbf{NLI Classifier} & \textbf{Accuracy} & \multicolumn{3}{c} {\textbf{Per class F1}} & \multicolumn{2}{c}{\textbf{Total}}\\ 
        
         &  & \multicolumn{1}{c}{T} &  \multicolumn{1}{c}{F} & \multicolumn{1}{c}{C} & \multicolumn{1}{c}{W-F1} & \multicolumn{1}{c}{M-F1} \\ \midrule
        \naive & 57.03 & 0.00 & 72.64 & 0.00 & 24.21 & 41.42 \\ 
        \gptt{} & 54.15 & 20.88 & 37.81 & 20.88 & 52.87 & 43.44 \\ 
        \gpttgold{} & 62.32 & 56.67 & 75.35 & 28.00 & 60.47&53.37 \\
        \robertaclaimnoev & 63.2 & 24.71 & 79.93 & 47.37 & 61.53 & 50.67 \\ 
        \robertaqtempplus & \underline{\textbf{\textcolor{gray}{65.25}}} &  \underline{\textbf{\textcolor{gray}{79.92}}} &  \underline{\textbf{\textcolor{gray}{47.76}}} &  \underline{\textbf{\textcolor{gray}{49.67}}} &  \underline{\textbf{\textcolor{gray}{66.56}}} &  \underline{\textbf{\textcolor{gray}{59.11}}} \\ 
        \robertagold & \textbf{69.66} & \textbf{56.86} & \textbf{82.92} & \textbf{48.79} & \textbf{69.79} & \textbf{62.85} \\ \bottomrule
    \end{tabularx}
    \\
    \caption{Quantitative analysis on evidence quality through claim verification performance comparison on \quantempplus.}
    \label{tab:data_quality_quantitative}
    \vspace{-3em}
\end{table*}

    \textbf{ \naive{}}: A Veracity Classifier that only predicts the majority class of the dataset. \robertaclaimnoev fine-tunes a NLI model to predict the veracity of the claim without evidence. 
    
    \textbf{ \gptt{}}: A few-shot learning model that given the claim and corresponding relevant evidence snippets collected using queries from \name{}, predicts the veracity of the claim similar to \cite{quantemp}. We use the \textit{gpt-3.5.-turbo} generative model for this purpose.
    
    \textbf{ \robertaqtempplus{}}: A pre-trained MNLI model fine-tuned to predict the veracity label of the claim given the claim and its corresponding relevant evidence snippets collected using queries from \name{}. \textbf{Please note} for this setting and \gptt{} above we do not perform retrieval from whole collection in open-domain setting as done in retrieval and fact-verification experiments discussed in Sections \ref{5.2} and \ref{5.3}. This setup is  to directly compare quality of evidence collected in \quantempplus{} with gold justification documents from manual fact-checkers.
    
    \textbf{  \robertagold{}}: A pre-trained MNLI model fine-tuned to predict the veracity label of the claim given the claim and its gold justification document authored by human fact-checker. This model serves as the upper bound that can be achieved with the pre-trained MNLI model for \quantempplus{}. 
 \gpttgold{} uses same setups as \robertagold{}, but with \textit{gpt-3.5.-turbo} for veracity prediction.

    \textbf{Fact-Verification:} We employ claim decomposition, followed by retrieval to fetch top-k evidences from our evidence collection in \quantempplus{}. These top-k evidences are then passed on to a NLI model along with the claim to perform verification. We employ Contriever as retrieval model afetr comparing several approaches like BM25, ANCE \cite{ance},Tas-b \cite{tas-b} on validation set.  We experiment with different values of $k=1,3,5,7,10$ for top-k evidence retrieval and observe k=3 to provide best NLI performance on validation set. We fine-tune all the NLI models including ones mentioned above using the Adam optimizer with a weight decay of \( 1\times e^{-5}\) and a learning rate of \(2 \times e^{-5}\). When LLMs are used for verification, we set temperature to 0.1 for reducing randomness in outputs.

    \textbf{Metrics}: For each of these settings, we evaluate the accuracy, per-class F1, macro-F1 (M-F1), and weighted-F1 (W-F1) scores.% to account for the class imbalance in the dataset.

   % \textbf{Hyperparameters}:

\section{Results}

\subsection{Quality of Evidence Collected through \name{}}
To answer \textbf{RQ1}, we evaluate the quality of the \quantempplus{} dataset both qualitatively and quantitatively.

To \textbf{quantitatively} measure the efficacy of our dataset, we fine-tune \robertamnlilarge{} with the training split of our dataset and evaluated its performance on our test split of claim and evidence pairs collecting using queries from \name{} based decomposition. From Table.\ref{tab:data_quality_quantitative} we see that the model trained on our dataset, \robertaqtempplus{}, performs much better than all the baselines on all metrics. It achieves relative gains of about 41\% over \naive{} and about 16\% over the \robertaclaimnoev{} classifier. We also see that the performance of \robertaqtempplus{} is closer to \robertagold{} (uses justification documents directly) indicating that we are very close to the verification upper bound. This demonstrates that queries from \emph{\name{} which emulates the search process in justification snippets from fact-checkers, are of better quality, resulting in better quality relevant evidences for each claim}. It is important to note \robertaqtempplus{} uses evidence without temporal leakage or gold justification leakage.

We also perform the qualitative evaluation of the queries from \name{} and resulting evidence collection in \quantempplus{} benchmark. Following the setup of ClaimDecomp \cite{claimdecomp}, we sample 50 representative claims across all categories of numerical claims to perform this analysis with help of 3 annotators who are researchers working on fact-checking. The inter-annotator agreement in presented in Table \ref{table:agreement}. We observe that the average precision of these queries is 90\%, meaning most of our queries are \textbf{relevant} to the claim. Additionally, we observe an average completeness score of \textbf{4.49} indicating queries from \name{} cover most of the aspects of the claim. The average redundancy score is \textbf{2.4} indicating mild redundancy due to our filtering process. Likewise for collected evidence snippets we observe an average completeness of \textbf{3.94} and redundancy of \textbf{2.9}.

\begin{table}[hbt!]
\centering
\begin{tabular}{lcc}
\toprule
\textbf{Metric} & \textbf{Queries} & \textbf{Evidence Snippets} \\
\midrule
\textbf{Completeness} & \cellcolor{green!10}0.61 & \cellcolor{green!10}0.66 \\
\textbf{Redundancy} & \cellcolor{yellow!10}0.4 & \cellcolor{yellow!10}0.45 \\
\textbf{Relevance} & \cellcolor{green!10}0.58 & \cellcolor{green!10}0.67 \\
\bottomrule
\end{tabular}
\caption{Inter-annotator agreement (Fleiss’) for qualitative analysis }%\quantempplus{}}
\label{table:agreement}
\vspace{-4em}
\end{table}
%\vspace{-2em}
\subsection{Retrieval Performance on \quantempplus{} }

\label{5.2}

To answer \textbf{RQ2}, we compare the retrieval performance of various claim decomposition approaches as shown in Table \ref{tab:ret_perf_temp}.  The retrieval is performed in an open-domain setup from entire evidence collection which also contains distractors. Over the top-k documents retrieved from the corpus using Contriever, we also add a \textbf{temporal filter} to retain only top-k documents that are published before the claim publication date. Though we already employ the temporal filter per-claim level when collecting evidence from the web, temporal leakage can still occur because evidences collected for other related claims in the benchmark maybe published at a later date. Hence, in an open-domain retrieval setup such evidences which are semantically related could be retrieved leading to temporal leakage. Prior benchmarks like AveriTec \cite{averitec} \textbf{do not account for this form of leakage in open-domain setup}. We observe that \name{} with temporal filter achieves the highest Recall@k for k=10,100 and also highest MRR which indicates higher ranking quality. This is primarily because \name{} generates queries by extracting the manual fact-checker's search process from justification documents.
\begin{table}[ht]
    \centering
    \scalebox{1.0}{
    \begin{tabular} { 
      lcccc}
        \toprule
        \textbf{Query mode} & \textbf{NDCG@10} & \textbf{Recall@10} & \textbf{Recall@100}  & \textbf{MRR} \\ \midrule
        \claimonly{} & 0.33 & 0.32 & 0.51 &  0.54 \\ \midrule
        \textbf{Decomposition}& & &  &\\ \midrule
        \oracleq{} & 0.57 & 0.54 & 0.82 & 0.68 \\ 
             \oracleq{}   (-temporal filter) & 0.54 & 0.51 & 0.82 & 0.651 \\
        \claimdecomp{} & 0.34 & 0.32 & 0.57  & 0.53 \\ 

                \claimdecomp{} (-temporal filter) & 0.31 & 0.30 & 0.56  & 0.49 \\ 
        \pgmfc{} & 0.32 & 0.31 & 0.52  &0.52\\ 

                \pgmfc{} (-temporal filter) & 0.30 & 0.29 & 0.52  &0.481\\ 
        \qgen{} & 0.33 & 0.31 & 0.56 & 0.57\\
          \qgen{} (-temporal filter) & 0.29 & 0.27 & 0.54 & 0.48\\\bottomrule
    \end{tabular}
    }
    \caption{Retrieval performance of different query planning methods using \combmaxn{} on \quantempplus{} with temporal filtering.}
    \label{tab:ret_perf_temp}
    \vspace{-2em}
\end{table}
While retrieval happens in a realistic setup one cannot assume access to justification documents as background information at test time to generate queries. Hence \name{} presents an upper bound on quality of queries and their impact on retrieval.  We observe that other decomposition approaches fall short in retrieval quality compared to \name{} as they try to decompose the claim without background information. Even \qgen{} which tries to distill \name{} queries from a LLM does not obtain optimal performance due to generation of queries based on claim alone without additional context. Hence, we hypothesize that iterative approaches that  claim decomposition and retrieval are necessary for optimal claim decomposition. We believe that by using additional information retrieved adaptively using intermediate queries/sub-claims to generate further queries  would help achieve better queries like \name{} and gains in retrieval performance. One \textbf{future direction}, is \emph{to develop new claim decomposition approaches  using queries from \name{} as ground truth}.

\vspace{-1em}

\subsection{Downstream Fact-verification Performance}
\label{5.3}
%\vspace{-1em}
While the above results indicate the performance of the query planning methods at retrieval, observing their corresponding downstream impact is essential as it represents the final outcome of the pipeline. Hence, to answer \textbf{RQ3} we evaluate the final performance of the veracity prediction component for each of the query planning methods, the results of which are shown in Table.\ref{table:downstream_nli}.

\emph{Superiority of decomposition-based methods.} Firstly, we observe that the performance of \robertaclaimnoev{} performs the worst as it relies only on surface level patterns in claim for veracity prediction without evidence. Secondly, we observe that the downstream performance provided by decomposition-based query planning methods, specifically that of \qgen{} is higher than that of \claimonly{}. 

\begin{table}[h!]
\centering
\begin{tabular}{lcccccc}
\toprule
\textbf{Query mode} & \multicolumn{6}{c}{\textbf{NLI performance}} \\
\cmidrule(lr){2-7}
& \textbf{Accuracy} & \textbf{F1-T}&\textbf{F1-F}&\textbf{F1-C}&\textbf{W-F1} & \textbf{M-F1} \\
\midrule
\robertaclaimnoev{} & 63.2 &24.71&79.93&47.37 & 61.53 & 50.67 \\
\oracle{} & 65.25&47.76&79.92&49.67 & 66.56 & 59.11 \\
\claimonly{} & 66.53&53.38&81.46&31.02 & 64.03 & \textbf{55.28} \\ \midrule
\textbf{Decomposition} & & &  \\ \midrule
\oracleq{} & 66.45&45.73&80.78&50.25 & 66.8 & \textbf{58.92}\textsuperscript{\textdagger} \\
\claimdecomp{} & 64.57&53.78&79.97&35.81 & 64.41 & 56.51 \\
\pgmfc{} & 65.67&51.38&81.19&37.15 & 65.73 & 56.57 \\
\qgen{} & 66.25&53.62&81.39&36.95 & 65.41 & \textbf{57.28}\textsuperscript{\textdagger} \\

\bottomrule
\end{tabular}
\caption{Fact Verification performance using different claim decomposition methods on \quantempplus{}.\textsuperscript{\textdagger} indicates statistical significance over \claimonly{} at 0.05 level. F1-T, F1-F, and F1-C - True, False, and Conflicting F1 scores.}
\label{table:downstream_nli}
    \vspace{-2em}
\end{table}

While \pgmfc{} and \claimdecomp{} which employ a larger model \gptt{} for decomposition are minimally better than \claimonly{}, \qgen{} which employs a much smaller model (Flan-T5-Large) provides a statistically significant gain of 3.6\% in the Macro-F1 score over \claimonly{} and is marginally better than \pgmfc{} and \claimdecomp{}. A comparison of the per-class F1 scores shows that decomposition-based query planning methods excel at claims that are of a conflicting nature. We observe relative gains of up to 20\% by \pgmfc{} compared to \claimonly{}. Claims whose veracity is of conflicting nature specifically require diverse perspectives to be retrieved as evidence since some parts of them may be true and other parts false. We observe that \claimonly{} fails to provide the correct downstream result as the evidence it retrieves depends only on the claim and is too homogeneous failing to capture multiple explicit and implicit aspects.

% \emph{Performance of \qgen{}.} Another interesting observation we see is that, despite \pgmfc{} and \claimdecomp{} using a larger LLM \gptt{} for decomposition, the downstream impact of \qgen{}'s claim decomposition method, which was developed by training a smaller model of Flan-T5 with our oracle queries from \quantempplus{}, is superior. The training of \qgen{} can also be seen as a form of knowledge distillation \cite{alpaca,vicuna,peng2023instruction,gu2024minillm}, as our oracle queries were generated by prompting the \gptt{} model with the claim and justification document as input.  

%Additionally, \cite{divide-conquer} has shown that problem decomposition tasks are easier to distill into small LLMs in QA tasks. We observe the same with decomposing claims into sub-queries, which aids in retrieving and verifying information.

\textbf{Discrepancy between retrieval and downstream performance:} An interesting observation that we see from comparing the results from retrieval in Table.\ref{tab:ret_perf_temp} and NLI from Table.\ref{table:downstream_nli} is that the gains in retrieval performance do not proportionally translate to downstream gains in NLI (fact-verification) performance. The retrieval performance of queries from \name{} provides significant relative gains %of %68.7\% at Recall@10 and 74\% at NDCG@10
over using \claimonly{} for retrieval. However, we see that this only translated to a relative gain of 6.6\% downstream. We observe that this may be \emph{primarily due to fundamental  numerical contextualization and understanding limitations in claim verification models and LLMs}. These models lack abilities to parse, contextualize, understand notion of ranges, thresholds and other numerical information in presence of other textual information.  Additionally, while retrieval systems optimize for end user consumption \cite{ret-enh-ml}, the retrieved evidences might not necessarily be useful for downstream tasks. Inspired by prior works in other tasks \cite{zhang-etal-2023-relevance} one could train the claim decomposition model by optimizing for downstream fact-verification performance in an end-end manner to reduce the observed gap in the performance of retrieval and downstream NLI.

\vspace{-1em}
\section{Conclusion}

In this work, we curate an open-domain benchmark for fact-checking  real-world numerical claims. We enhance the evidence collection for realistic open-domain fact-verification setup through a claim decomposition mechanism emulating a manual fact-checkers search process followed by automated filtering mechanisms to prevent temporal leakage and gold justification leakage. On evaluation of quality of evidence snippets, we observe improvements compared to upper bound performance and prior works like QuanTemp. Our benchmark paves way for clear future directions enabling to develop better claim decomposition mechanisms. We also introduce the retrieval-verifier gap enabling future research directions in this area.

 \bibliographystyle{splncs04}
 \bibliography{references}

\end{document}